\documentclass[conference]{IEEEtran}
\usepackage{cite}
\usepackage{amsmath,amssymb,amsfonts}
\usepackage{algorithmic}
\usepackage{graphicx}
\usepackage{textcomp}
\usepackage{xcolor}

\usepackage{diagbox}
\usepackage[caption=false]{subfig}
\usepackage{tabularx}
\usepackage[]{mdframed}
\usepackage{booktabs}
\usepackage{makecell} %
\usepackage{threeparttable} %
\usepackage{enumitem}
\usepackage{url}
\usepackage{tcolorbox}
\usepackage[
colorlinks=true,
    linkcolor=blue,
    citecolor=blue,
    urlcolor=blue,
    filecolor=blue,
    anchorcolor=blue,
    pdfborder={0 0 0},
    bookmarks=true,
    bookmarksnumbered=true,
    bookmarksopen=true,
    pdfstartview=FitH,
    pdfpagemode=UseOutlines
]{hyperref}
\usepackage{listings}
\usepackage{pifont} %

\hypersetup{
    pdftitle={Legal Compliance Evaluation of Smart Contracts Generated by Large Language Models},
    pdfauthor={Chanuka Wijayakoon, Hai Dong, H.M.N. Dilum Bandara, Zahir Tari, Anurag Soin},
    pdfkeywords={large language models, smart contracts, legal compliance, measurement, software metrics},
}

\newtcolorbox{outerbox}[2][]{
    colback=blue!5!white,
    colframe=blue!75!black,
    coltitle=white,
    colbacktitle=blue!85!black,
    fonttitle=\bfseries,
    title={#2},
    halign=left,
    left=1mm,
    right=1mm,
    subtitle style={
        boxrule=0.4pt,
        colback=blue!15!white,
        colupper=blue!15!black,
    },
    #1
}

\newtcolorbox{innerbox}[1][]{
    colback=white,
    colframe=blue!25,
    fonttitle=\bfseries,
    title=#1,
    colbacktitle=gray!20,
    attach title to upper=\par,
    coltitle=black,
    halign=left,
    boxsep=1mm, %
    fontupper=\footnotesize
}

\def\BibTeX{{\rm B\kern-.05em{\sc i\kern-.025em b}\kern-.08em
    T\kern-.1667em\lower.7ex\hbox{E}\kern-.125emX}}
\begin{document}

\title{Legal Compliance Evaluation of Smart Contracts Generated by Large Language Models
\thanks{We gratefully acknowledge funding from the Digital Finance CRC which is supported by the Cooperative Research Centres program, an Australian Government initiative. We also thank the Australia and New Zealand Bank for their support and guidance during this research.}
}

\makeatletter
\newcommand{\linebreakand}{%
  \end{@IEEEauthorhalign}
  \hfill\mbox{}\par
  \mbox{}\hfill\begin{@IEEEauthorhalign}
}
\makeatother

\author{
    \centering

    \IEEEauthorblockN{Chanuka Wijayakoon\IEEEauthorrefmark{1}, 
                  Hai Dong\IEEEauthorrefmark{1}, 
                  H.M.N. Dilum Bandara\IEEEauthorrefmark{2}, 
                  Zahir Tari\IEEEauthorrefmark{1}, 
                  Anurag Soin\IEEEauthorrefmark{3}
                  }
\IEEEauthorblockA{
\IEEEauthorrefmark{1}School of Computing Technologies, RMIT University, Melbourne, Australia\\
                  chanuka.wijayakoon@student.rmit.edu.au, hai.dong@rmit.edu.au, zahir.tari@rmit.edu.au.
                  }
\IEEEauthorblockA{
\IEEEauthorrefmark{2}CSIRO's Data61, Sydney, Australia\\
dilum.bandara@csiro.au.
}
\IEEEauthorblockA{
\IEEEauthorrefmark{3}University of Technology, Sydney, Australia\\
                  anurag.soin@student.uts.edu.au.
                  }
}
\maketitle

\begin{abstract}
Smart contracts can implement and automate parts of legal contracts, but ensuring their legal compliance remains challenging. Existing approaches such as formal specification, verification, and model-based development require expertise in both legal and software development domains, as well as extensive manual effort. Given the recent advances of Large Language Models (LLMs) in code generation, we investigate their ability to generate legally compliant smart contracts directly from natural language legal contracts, addressing these challenges. We propose a novel suite of metrics to quantify legal compliance based on modeling both legal and smart contracts as processes and comparing their behaviors. We select four LLMs, generate 20 smart contracts based on five legal contracts, and analyze their legal compliance. We find that while all LLMs generate syntactically correct code, there is significant variance in their legal compliance with larger models generally showing higher levels of compliance. We also evaluate the proposed metrics against properties of software metrics, showing they provide fine-grained distinctions, enable nuanced comparisons, and are applicable across domains for code from any source, LLM or developer. Our results suggest that LLMs can assist in generating starter code for legally compliant smart contracts with strict reviews, and the proposed metrics provide a foundation for automated and self-improving development workflows.
\end{abstract}


\section{Introduction}
\label{sec:intro}

Blockchains are increasingly used for multi-party business processes~\cite{Mendling2018}, with users making direct agreements via immutable programs known as \emph{smart contracts}~\cite{mavridou2019verisolid}. With the tokenized asset market projected to reach US \$16.1 trillion by 2030~\cite{kumar2022relevance}, businesses are rapidly developing smart contract-based services on blockchain platforms. To avoid significant legal, financial, and reputational risks, these smart contracts must comply with legal constraints that arise from legal contracts between stakeholders~\cite{dwivedi2021legal,idelberger2016evaluation} and regulatory frameworks in the operating jurisdiction~\cite{auer2019embedded}.

Manually implementing legally compliant code is a time-consuming and error-prone process, especially for complex multi-party agreements. With recent advancements in large language models (LLMs) such as GPT-4, there has been a growing interest in using generative models to write code. While LLMs can produce syntactically correct code~\cite{austin2021programsynthesislargelanguage}, generating smart contracts from natural-language legal agreements faces questions regarding legal compliance. Existing LLM evaluations largely focus on syntactic quality~\cite{chen2021evaluating} and basic semantic accuracy~\cite{liu_chatgptCorrectness}. However,, legal compliance analysis requires complex semantic interpretations and has garnered less attention, while LLMs' legal reasoning capabilities show high variability~\cite{guha2023legalbench}. Furthermore, traditional source-code metrics, such as Cyclomatic Complexity~\cite{mccabe1976complexity}, fail to capture crucial legal compliance aspects such as rights, obligations, and temporal constraints.

\begin{figure}[t!]
    \centering
    \includegraphics[width=\columnwidth]{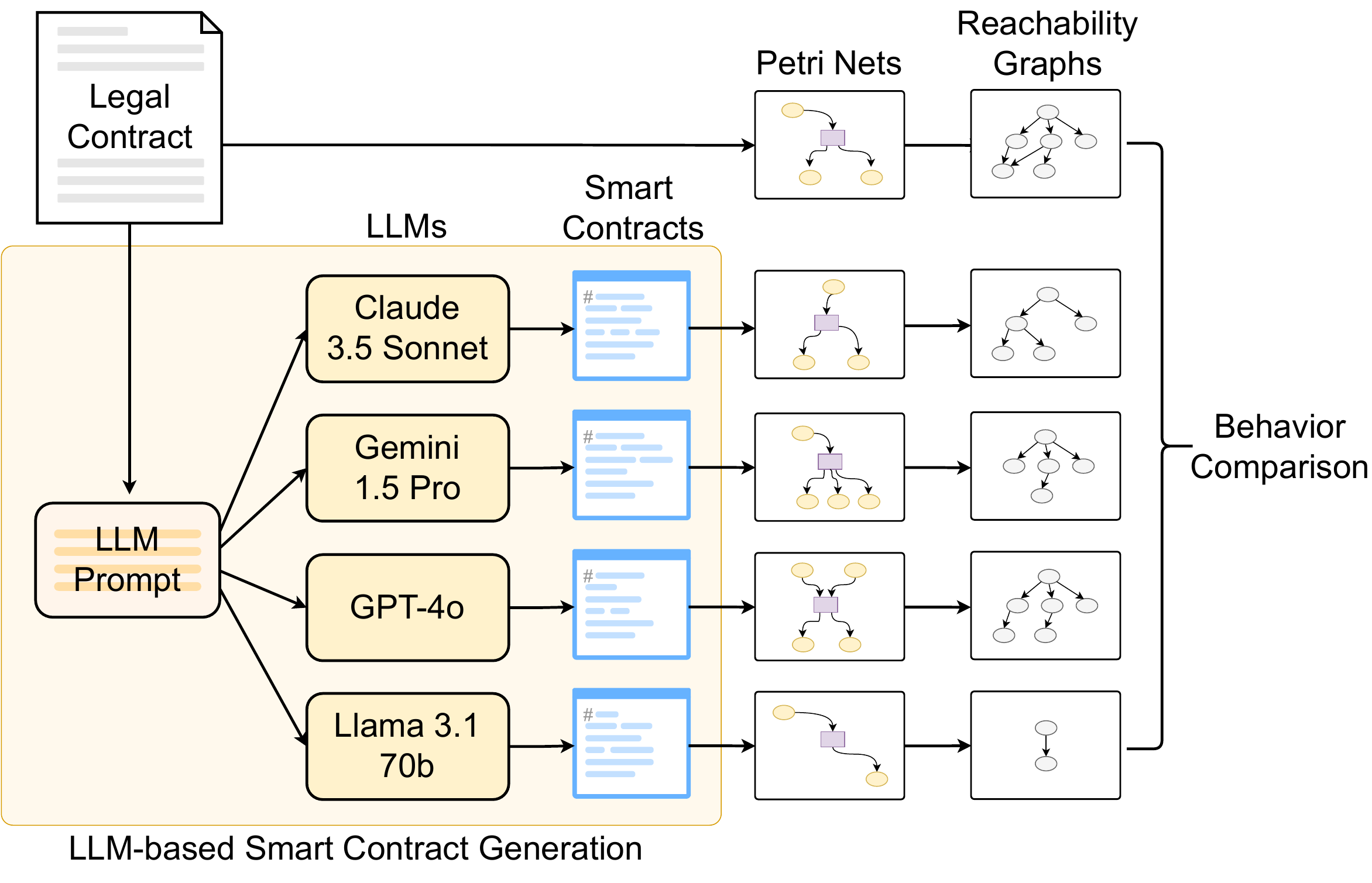}
    \vspace{-20pt}
    \caption{Workflow of Petri-net-based legal and smart contract behavior analysis.}
    \label{fig:workflow_summary}
    \vspace{-15pt}
\end{figure}

To address this gap, we propose a systematic evaluation of the legal compliance of smart contracts generated by leading LLMs, as shown in Fig.~\ref{fig:workflow_summary}. We focus on the alignment between the behaviors captured in a natural-language legal contract and those expressed in the corresponding smart contract, and use it as a proxy to assess LLMs' capability to generate legally compliant code. To facilitate this comparison, we adopt a contracts-as-processes view~\cite{daskalopulu2000}, modeling both legal and smart contracts as Petri Nets (PNs)~\cite{Petri_Reisig_2008} to analyze their behaviors. 

We introduce a novel suite of metrics for legal compliance: 1) \emph{fitness}, measuring how many behaviors from the legal contract are captured in the smart contract; 2) \emph{precision}, assessing how many behaviors in the smart contract are relevant to the legal contract; and 3) \emph{functional equivalence score} (FES), providing a flexible assessment of the alignment between the two contracts. These metrics allow us to quantify the degree of legal compliance for each LLM-generated smart contract, offering a nuanced understanding of how different LLMs perform. As the behavior analysis is independent of how a smart contract is conceived, these metrics can also be applied to any smart contract, from an LLM or a developer.

In this paper, we present three primary contributions:
\begin{enumerate}
    \item A suite of software metrics based on behavior analysis to quantify the legal compliance of source code.
    \item A PN-based method to systematically calculate the metrics to determine legal compliance of LLM-generated smart contracts against natural-language legal contracts.
    \item A comparative evaluation of smart contracts generated by GPT-4o, Gemini 1.5 Pro, Llama 3.1 70b, and Claude 3.5 Sonnet, highlighting their strengths and weaknesses in achieving legal compliance.
\end{enumerate}

The paper is structured as follows: Sec.~\ref{sec:background} outlines background information for our analysis. Sec.~\ref{sec:methodology} presents the research methodology, with our metrics suite in Sec.~\ref{sec:metrics_and_application}. Sec.~\ref{sec:evaluation} outlines the experimental setup and presents the results of our smart contract legal compliance evaluation. We discuss our findings and address potential threats to validity in Sec. \ref{sec:discussion}. Sec.~\ref{sec:literature} reviews related work. Concluding remarks and future directions are in Sec.~\ref{sec:conclusion}.

\section{Preliminaries}
\label{sec:background}

We assess smart contract legal compliance by comparing their behaviors against their base legal contracts. Our approach models both contract types as PNs, providing a formal framework for behavioral comparison. This section outlines the theoretical foundation of our methodology.

\subsection{Legal Powers and Obligations}

Legal contracts are structured around fundamental legal concepts such as rights, duties, powers, and obligations~\cite{hohfeld}. These concepts are the foundation for the behaviors that smart contracts need to implement. LLMs, when generating smart contracts, must correctly interpret these legal concepts from the natural-language contract and encode them in code. 

In this paper, we follow the Symboleo framework~\cite{parvizimosaed2022symboleo}, a formal specification language designed for modeling legal contracts, in choosing legal concepts for analysis: power and obligation. A \emph{power} is a legal right that enables one party to change the legal positions of another party. An \emph{obligation} is a duty that requires one party to take specific actions when certain conditions are met. 

\subsection{Petri Nets}
\label{subsec:Prelim_PN}

PNs~\cite{Petri_Reisig_2008} provide a formal graphical representation to model concurrent systems. They consist of \emph{places} (ovals) representing resource status or conditions, \emph{transitions} (rectangles) denoting actions or events, \emph{tokens} to indicate the current state of the system, and \emph{arcs} (directed edges) showing the direction of workflows and token movement. Arcs only connect places to transitions or transitions to places.

A token in a place indicates an active resource status or a true condition. A transition is enabled if all its incoming places (called the \emph{preset}) have tokens. Upon firing, it consumes a token from each incoming place and puts one token each in its outgoing places (called the \emph{postset}). An arc with arrow heads on both ends denotes that an associated transition both consumes and produces a token at the connected place. There is also a special \emph{inhibitor arcs} type that inhibits a transition from firing if the connected place has a token. All the possible states of the system represented by different token configurations are captured by the PN's reachability graph (RG). We refer readers to Gehlot~\cite{Gehlot2019} for a more detailed introduction to PNs.

We choose PNs as our modeling formalism for several key reasons. First, their support for concurrent execution and state transitions naturally maps to legal~\cite{bons1995modelling} and smart contract~\cite{duoSCformalAnalysisCPN2020} behaviors, where multiple parties can independently initiate contract actions. Second, powers and obligations in legal contracts can be directly represented with places and transitions in PNs~\cite{bons1995modelling}, allowing intuitive modeling of legal positions and their state changes. Third, PNs have formal semantics that enable rigorous comparison of behaviors through reachability analysis~\cite{Petri_Reisig_2008}. Lastly, PNs' visual nature helps bridge communication between legal and technical teams---a crucial advantage when evaluating legal compliance.

\subsection{Events, Behaviors, and Legal Relevance}

An \emph{event} is an immutable occurrence that happens at a particular time, and a \emph{behavior} is a sequence of events that occurs in a given contract. Events are legally relevant if they affect the legal positions of contract parties. As an example, fulfilling a payment obligation would be a legally relevant event in a sales contract. In PNs, each path in a RG represents a behavior. We consider two forms of event equivalence when comparing behaviors:
\begin{enumerate}
    \item \emph{Strict event equivalence}: each event in the ground truth behavior must have one or more events that result in a functionally equivalent outcome, in the same order as the ground truth. This means that for behaviors $B_p = <e_{p_1}, e_{p_2}, ... e_{p_m}>$ from ground truth system and $B_q = <e_{q_1}, e_{q_2}, ...e_{q_n}>$ from candidate system, each $e_{p_i}$ maps to one or more events from the sequence $e_{q}$, while preserving the order of events. There can be extra events at the end of a candidate behavior after accounting for all events from the ground truth behavior unless they cause a legal state deviation.
    \item \emph{Non-strict event equivalence}: allows an event in a behavior of the ground truth to have no equivalent event(s) in the candidate system.
\end{enumerate}

Building on these concepts, we next present how we derive and calculate the proposed compliance metrics.

\section{Methodology}
\label{sec:methodology}

Fig.~\ref{fig:workflow_summary} shows our approach to analyze LLM-generated smart contracts' legal compliance. We model legal contracts as PNs using \emph{CPN Tools}~\cite{cpntools} and \emph{CPN IDE}~\cite{verbeek2021cpnide}, then generate corresponding smart contracts using four LLMs and model them also as PNs. We derive RGs from these PNs to analyze states and execution paths. The PNs and RGs reflect any smart contract errors, which our metrics suite aims to capture. We compare these graphs against the legal contract's RG as ground truth. The source contracts, smart contracts, and PNs are available online{\footnote{\url{https://doi.org/10.5281/zenodo.14074462}}}.

\subsection{Petri-Net Modeling}
Following Symboleo~\cite{parvizimosaed2022symboleo}, we consider only \emph{power} and \emph{obligation} legal positions as they are monitorable. To analyze the state changes of these legal positions, we construct PNs from legal and smart contracts. For the legal contract, we consider the point at which a fresh instance of the contract may be initialized as the starting state of its PN. Then the logical execution of legal positions is used to develop the rest of the PN. Similarly, each smart contract is initialized in a state mirroring that of the legal contract's initial state. We use basic PNs without token differentiation, which limits the system to single instances of contract parties or resources.

We use a simplified version of the USDC stablecoin Terms of Service contract~\cite{USDCToS} to illustrate our analysis. This contract, called GCDC ToS (Generic Currency Digital Coin Terms of Service) from here onwards, has no references to USDC, its issuing company, or related entities, so that LLMs are not unduly influenced by their preexisting knowledge of USDC. The contract, shown in Fig.~\ref{fig:gcdc_contract}, is between the GCDC issuing company (XYZ) and GCDC platform user (User).

Certain workflows were omitted from the full GCDC contract for clarity and to maintain a manageable scope. For example, blacklisting, freezing, and terminating accounts lead to similar outcomes. Thus, blacklisting is selected as the representative workflow excluding others.

\begin{figure}[t]
    \centering
    \begin{tcolorbox}[
      colback=white, %
      colframe=black!60, %
      coltitle=black, %
      fonttitle=\bfseries\footnotesize, %
      colbacktitle=gray!20, %
      fontupper=\footnotesize, %
      sharp corners, %
      boxrule=0.5pt, %
      toprule=2pt, %
      bottomrule=2pt, %
    ]
    \setlength{\leftmargini}{0.25em}
    \begin{itemize}
    \item By holding or using GCDC, or using any of the GCDC Services, you (“you,” “your,” or “User”) agree that you have read, understood and accept all of the terms and conditions contained in these Terms, as well as our (``XYZ'') Privacy Policy, Cookie Policy and E-Sign Consent, and you acknowledge and agree that you will be bound by these terms and policies.

    \item As you have agreed to, and are subject to, the XYZ Mint account User Agreement, XYZ makes available the following GCDC-related Services as defined in the XYZ Mint account User Agreement:
    \begin{itemize}
        \item[\rm(i)] issue GCDC for USD from XYZ, 
        \item[\rm(ii)] redeem GCDC for USD from XYZ, and 
        \item[\rm(iii)] send and receive GCDC to and/or from XYZ Mint accounts (collectively, the “GCDC Services”).
    \end{itemize}

    \item GCDC is issued and redeemed in accordance with XYZ's blacklisting policy. XYZ reserves the right to block the transfer of GCDC to and from an address on chain as permitted under the blacklisting policy.

    \item We reserve the right to (i) change, suspend, or discontinue any aspect of the GCDC Services at any time, including hours of operation or availability of any feature, without notice and without liability and (ii) decline to process any issuance  or redemption without prior notice and may limit or suspend your use of one or more GCDC Services at any time, in our sole discretion.
\end{itemize}
    \end{tcolorbox}
    \vspace{-5pt}
    \caption{Excerpt from the GCDC ToS contract, adapted from~\cite{USDCToS}.}
    \label{fig:gcdc_contract}
    \vspace{-15pt}
\end{figure}

\begin{table}[tb]
    \caption{Legal positions of the GCDC contract}
    \vspace{-3pt}
    \footnotesize
    \centering
    \begin{tabularx}{\columnwidth}{l X}
        \toprule
            \textbf{Identifier} & \textbf{Legal Position} \\
            \midrule
            $P_{ReqIssue}$ & User's power to request XYZ to issue GCDC. \\
            $P_{ReqRedeem}$ & User's power to request XYZ to redeem GCDC. \\
            $P_{SendGCDC}$ & User's power to send GCDC. \\
            $P_{IssueGCDC}$ & XYZ's power to issue GCDC. \\
            $P_{RedeemGCDC}$ & XYZ's power to redeem GCDC. \\
            $P_{Blacklist}$ & XYZ's power to blacklist an address. \\
            $P_{Pause}$ & XYZ's power to pause smart contract functionality. \\
        \bottomrule
    \end{tabularx}
    \label{tab:legal_positions}
    \vspace{-5pt}
\end{table}

\begin{figure}[b!]
        \centering
        \subfloat[Petri net of GCDC contract.]{    
            \includegraphics[width=0.99\columnwidth, height=6.5cm]{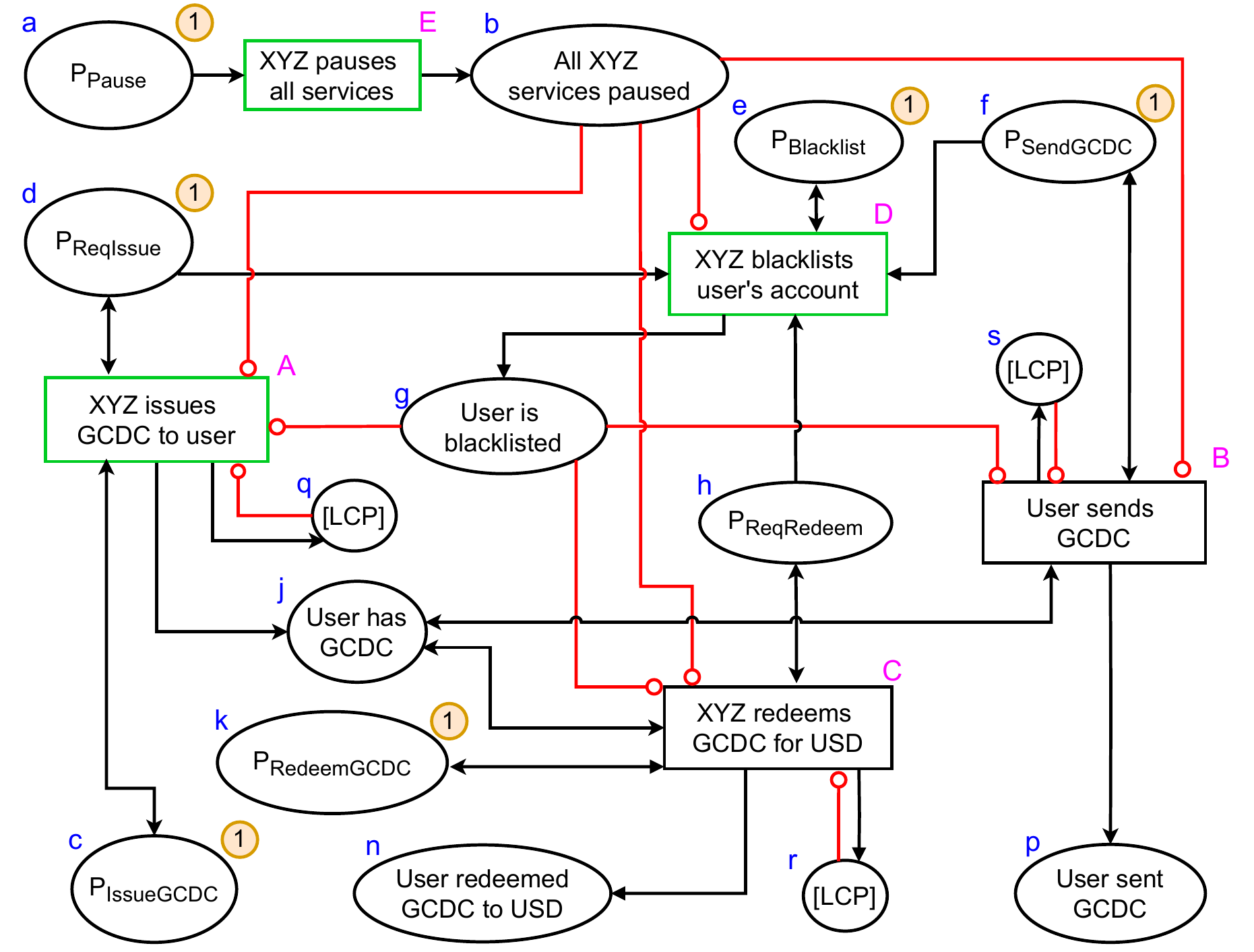}
            \label{fig:GCDC_PN}
        } \\
        \subfloat[Reachability graph of GCDC contract.]{
            \includegraphics[width=0.99\columnwidth, height=5.5cm]{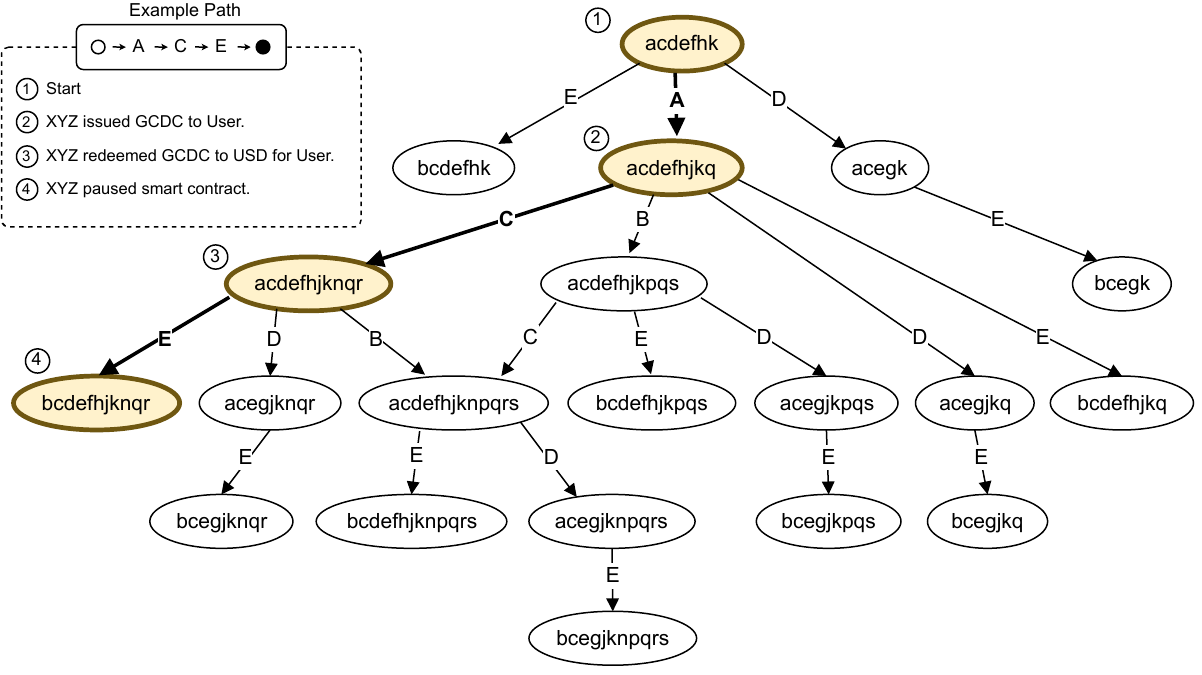}
            \label{fig:GCDC_RG}
        }
    \caption{Petri net and reachability graph of the GCDC Terms of Service.}
    \label{fig:GCDC_PN_RG}
    \vspace{-15pt}
\end{figure}

\subsubsection{Modeling Legal Contracts as PNs}
Here, we first identify the legal positions in the contract, as shown for the GCDC ToS contract in Table \ref{tab:legal_positions}. Next, a PN is drawn considering the logical execution of the contract terms, resulting in models similar to Fig.~\ref{fig:GCDC_PN}. We worked with two lawyers to minimize errors and ensure accurate representation of the underlying legal conditions, as their models are used as the ground truth. Discrepancies were resolved through discussion until lawyers reached consensus. We do not include behaviors that may be implied beyond the explicit functional scope of the base legal contract. We used CPN IDE's simulation tool to ensure there are no dead transitions. RGs, similar to Fig.~\ref{fig:GCDC_RG} for GCDC ToS, are derived using CPN Tools. We manually analyze each path to ensure they adhere to the stipulated conditions.

\subsubsection{Modeling Smart Contracts as PNs}
We systematically transform smart contracts to PNs by examining logical control flow and variable changes, following a four step process. First, we identify contract ownership and initial states with contextual variable interpretation. Second, we analyze execution paths and branching logic through code review. Next, we map temporal conditions and ordering constraints to PN structures, with inhibitor arcs for sequential ordering. Finally, we handle legally irrelevant operational details, such as numeric arguments, file paths, and subjective multipliers, through assumptions. Empty methods representing trackable off-chain events are treated as functional components, and missing implementations in state-critical methods are flagged as bugs. Temporal conditions use return arcs unless the consuming transition halts the contract, and they are modeled as independent behaviors only when producing operational effects on contract state.

\emph{Loop control places (LCP):} Some transitions both consume and produce a token from and to the same place, such as transition $A$ and place $d$, $P_{ReqIssue}$ in Fig.~\ref{fig:GCDC_PN}. These transition-place pairs create self-loops that can fire repeatedly but yield no new information after first execution. To limit them, we introduced a loop-control place in their postset to flag the initial firing, connecting it to the transition with an inhibitor arc, similar to place $q$. These flag places are initially empty, but receive a token at the first firing of the associated transition, preventing subsequent firings due to inhibitor arcs.

\subsection{Smart Contract Petri Nets Analysis}
Taking the legal contract's RG as the ground truth, we compare each of its paths against those in the smart contracts' RGs. Because both contract forms are initialized in equivalent states, their RGs are derived from equivalent initial markings. Each path in the legal contract's RG, as seen in Fig.~\ref{fig:GCDC_RG} for the GCDC contract, represents a distinct sequence of events, i.e., a behavior, through which legal positions change their states. Considering these behaviors, we calculate the proposed compliance metrics.

\section{Metrics for Legal Compliance}
\label{sec:metrics_and_application}

We propose a metric suite designed to measure legal compliance between a legal contract (serving as the ground truth) and its smart contract implementation, based on their exhibited behavior sets. While the legal contract exists in natural language, smart contracts are written in imperative programming languages and may implement the same functionality differently. Though we focus on LLM-generated smart contracts, these metrics are applicable to any smart contract implementation, generated automatically or coded manually. While we choose PNs for reasons stated in Sec.~\ref{subsec:Prelim_PN}, behavior modeling may be done with other suitable formalisms.

\begin{figure}[t!]
    \centering
    \subfloat{
        \includegraphics[height=5.1cm]{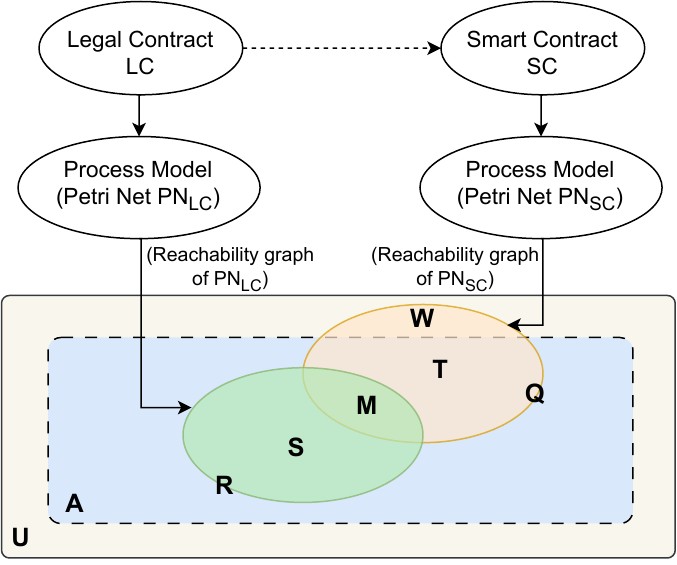}
        \label{fig:behaviors_set_diagram}
    } \\
    \vspace{-5pt}
    \subfloat{
        \begin{footnotesize}
        \begin{tcolorbox}[
            colback=white,
            colframe=black!50,
            arc=2mm,
            boxrule=0.5pt,
            left=1mm,
            right=1mm,
            top=1mm,
            bottom=1mm,
            width=0.95\columnwidth
        ]
        \begin{tabularx}{\linewidth}{@{\hspace{2mm}}cX@{\hspace{2mm}}}
            \textbf{U} & All possible behaviors\\
            \textbf{A} & All behaviors (both explicit and implicit through logical inference) enabled by the legal contract\\
            \textbf{R} & Behaviors in the legal contract's Petri net\\
            \textbf{Q} & Behaviors in the smart contract's Petri net\\
            \textbf{M} & Behaviors in both the legal and smart contracts' Petri nets\\
            \textbf{S} & Behaviors only in the legal contract's Petri net\\
            \textbf{T} & Behaviors in the smart contract's Petri net that are implicitly among those of the legal contract\\
            \textbf{W} & Behaviors in the smart contract's Petri net that are not in that of the legal contract, even implicitly
        \end{tabularx}
        \end{tcolorbox}
        \label{tab:behavior_sets_table}
        \end{footnotesize}
    }
    \vspace{-5pt}
    \caption{Possible behaviors of a legal contract and its smart contract.}
    \label{fig:behaviors_sets}
    \vspace{-15pt}
\end{figure}

As shown in Fig.~\ref{fig:behaviors_sets}, a legal contract exhibits a range of behaviors (set $A$) comprising of explicitly defined terms (set $R$) and logically inferred behaviors, with ambiguous boundaries due to interpretation challenges\cite{parvizimosaed2022symboleo}. A smart contract implementation may capture some legal behaviors (set $M$) while introducing behaviors that deviate from but do not violate the legal contract (set $T$), or are entirely unrelated (set $W$). Our proposed metrics account for these varying behavior types, recognizing that compliance requirements may range from implementing a single legal behavior to capturing all specified behaviors.

We propose three metrics to evaluate the legal compliance of smart contracts: fitness, precision, and functional equivalence score (FES). All of them expect \textit{legal equivalence}---the equivalence of states of powers and obligations---at the end of each behavior. The goal is to compare the outputs from each LLM and determine which LLM produces a smart contract most aligned with the legal ground truth.

\subsection{Fitness}
\label{sec:fitness}

The \emph{fitness} metric measures how closely the behaviors of a smart contract (SC) match those of the legal contract (LC). Fitness is a strict metric that evaluates the completeness of the smart contract’s inclusion of behaviors from the legal contract, based on strict event equivalence.

For each behavior $B_p$ in the legal contract, we check if there exists a functionally equivalent behavior $B_q$ in the smart contract, where every event in $B_p$ has a corresponding event in the same order in $B_q$. We identify one exception to this requirement: the passing of time, such as deadline expiry, may not be captured explicitly in a smart contract. Thus, temporal events (and associated changes of legal powers and obligations) are exempt from the strict event equivalence requirement given that they may be realized without any code execution. Event ordering must still be preserved, and the contracts must reach legally equivalent states at the end. Additionally, when multiple behaviors in a smart contract map to a behavior in the legal contract, they are counted as one.

The \emph{fitness} $F(SC, LC)$ of a smart contract to the legal contract is defined as the ratio of behaviors that strictly match between the contracts to the total number of behaviors in the legal contract. We observe its similarity to \emph{recall} in information retrieval, which is defined as the number of relevant instances retrieved from all relevant instances. With behavior sets (see Fig.~\ref{fig:behaviors_sets}) $R$ from the legal contract and $Q$ from the smart contract, this metric is formalized as follows:

\begin{equation}
F(SC, LC) = \frac{| R \cap Q |}{|R|}
\end{equation}

Fitness allows us to compare the accuracy of smart contracts generated by different LLMs. For example, if the fitness of a smart contract generated by LLMs $A$ and $B$ are 0.7 and 0.5, respectively, $A$ better captures the set of expected behaviors from the legal contract.

\subsection{Precision}
\label{sec:precision}

\emph{Precision} measures how much extra behavior a smart contract introduces beyond those in the legal contract. It is a non-strict metric (denoted by * in behavior sets) that allows for partial event inclusion and focuses on the degree of over-approximation. The two contract forms must still reach legally equivalent states at the end of corresponding behaviors.

For each behavior in the smart contract, we check if it contains events that are not strictly necessary or are absent in the legal contract. This allows identifying behaviors in the smart contract that deviate from or go beyond the required legal behaviors. Precision is defined as the ratio of behaviors that match between the legal and smart contracts to the total number of behaviors in the smart contract:

\begin{equation}
Pr(SC, LC) = \frac{| (R \cap Q)^{*} |}{|Q|}
\end{equation}

Precision helps compare how well each LLM-generates a smart contract that remains close to the specified behaviors without introducing unnecessary or incorrect behaviors. For instance, if the precision of contracts from LLM $A$ and $B$ are 0.8 and 0.6, the former has fewer extraneous behaviors than the latter.

\subsection{Functional Equivalence Score (FES)}
\label{sec:fes}

\emph{FES} measures the overall functional similarity between the legal contract and the smart contract, relaxing the strict event equivalence requirement of the fitness metric. It compares whether at the end of corresponding behaviors, both the legal and smart contracts reach legally equivalent states, even if the event sequences differ. It is defined as the ratio of behaviors in the smart contract that are functionally equivalent to those in the legal contract, regardless of strict event matching:

\begin{equation}
FES(SC, LC) = \frac{| (R \cap Q)^{*} |}{|R|}
\end{equation}

FES provides a flexible measure of how well the LLM-generated smart contract adheres to the legal contract’s intent, even with implementation differences. For example, a contract from LLM $A$ may have a perfect FES of 1, meaning every behavior from the legal contract is functionally captured, even if the event sequences differ.

\section{Evaluation}
\label{sec:evaluation}

\subsection{Experimental Setup}
\subsubsection{Legal Contract Selection}
We adopt the natural-language legal contracts used with the Symboleo framework~\cite{parvizimosaed2022symboleo} for our analysis. These are selected as they are monitorable, generalized versions of commonly found real-world contracts, representative of what may be encountered in business activities. Specifically, we selected four contracts from Symboleo dataset with varying numbers of legal positions and the GCDC ToS contract to evaluate different levels of complexity with the proposed metrics suite. An overview of the selected contracts is shown in Table~\ref{tab:lc_overview}, and they all involve at least two parties.

\begin{table}[t!]
    \centering
    \begin{threeparttable}
        \caption{Overview of legal contracts}
        \label{tab:lc_overview}
        \begin{tabularx}{\linewidth}{l c c c}
            \toprule
            \textbf{Contract Name} & \textbf{Obligations} & \textbf{Powers} & \textbf{Behaviors}\tnote{*} \\
            \midrule
            GCDC Terms of Service         & 0 & 7 & 12 \\
            Meat sale                              & 3 & 3 & 12 \\
            Pizza delivery                         & 3 & 2 & 3 \\
            Shipper-Carrier                        & 3 & 0 & 3 \\
            Transactive energy                     & 3 & 3 & 24 \\
            \bottomrule
        \end{tabularx}
        \begin{tablenotes}
            \footnotesize
            \item * Behaviors in the reachability graph of the legal contract's PN.
        \end{tablenotes}
    \end{threeparttable}
    \vspace{-7pt}
\end{table}

\subsubsection{LLM Selection and Configuration} We selected four LLMs, GPT-4o~\cite{openai2024gpt4ocard}, Gemini 1.5 Pro~\cite{gemini15}, Claude 3.5 Sonnet~\cite{claude35}, and Llama 3.1 70b Instruct~\cite{dubey2024llama},  based on their architectural diversity (proprietary and open-source), demonstrated performance in code generation and reasoning, and enterprise viability. GPT-4o, Claude 3.5 Sonnet, and Gemini 1.5 Pro represent state-of-the-art reasoning capabilities~\cite{claude2024ModelCardAddendum}, and Llama 3.1 70b represents leading open-source alternatives~\cite{dubey2024llama}, allowing for a comprehensive evaluation. For every model, we set the context length to $131,072$ tokens, temperature to $0.8$, top-K to $40$, and top-P to $0.9$. Maximum output tokens value is set to the maximum of each model. A summary of the models is shown in Table~\ref{tab:language_models}.

\begin{table}[t]
    \centering
    \begin{threeparttable}
        \caption{Large anguage models and their characteristics}
        \begin{tabularx}{\linewidth}{@{}llrrl@{}}
        \toprule
        \textbf{Model} & \textbf{Knowledge} & \textbf{Context} & \textbf{Max} & \textbf{Open} \\
        & \textbf{Cutoff} & \textbf{Window} & \textbf{Output} & \textbf{Source}\\
        \midrule
        Claude 3.5 Sonnet\tnote{*} & Apr, 2024 & 200K & 8K & No \\
        Gemini 1.5 Pro\tnote{†} & Oct, 2023 & 2M & 8K & No \\
        GPT-4o\tnote{‡} & Oct, 2023 & 128K & 16K & No \\
        Llama 3.1 70b Instruct\tnote{§} & Dec, 2023 & 128K & 4K & Yes \\
        \bottomrule
        \end{tabularx}
        \begin{tablenotes}\footnotesize
        \item * --claude-3-5-sonnet-20241022, † --gemini-1.5-pro-002, ‡ --
        gpt-4o-2024-08-06, § --llama-3.1-70b-instruct. K = 1,024 tokens, M = 1,024K tokens
        \end{tablenotes}
        \label{tab:language_models}
    \end{threeparttable}
    \vspace{-10pt}
\end{table}

\subsubsection{Prompting Strategy} We adopt a structured zero-shot prompting strategy based on the Plan-and-Solve (PS) approach~\cite{wang2023planandsolveprompting}. Through six stepwise prompts (see Fig.~\ref{fig:llm_final_prompt}), the LLM systematically extracts key contract elements, i.e., parties, legal positions, triggers, time frames, and dispute mechanisms, before generating the smart contract code. Initial attempts using zero-shot Chain-of-Thought~\cite{wei2022chain} and two-step prompting faced timeout and information capture issues. Through iterative refinement and GPT-4o's optimization, we developed our final PS-styled prompts that resolved these limitations.
\begin{table*}[tb!]
\centering
\begin{threeparttable}
\caption{Legal compliance metrics}
\label{tab:legal_compliance_metrics}
\begin{tabularx}{\textwidth}{@{}p{2.7cm} *{12}{>{\raggedleft\arraybackslash}X}@{}}
\toprule
& \multicolumn{4}{c}{FES}          & \multicolumn{4}{c}{Fitness}       & \multicolumn{4}{c}{Precision}     \\
\cmidrule(lr){2-5} \cmidrule(lr){6-9} \cmidrule(lr){10-13}
Contract        & CS & GP & GPT & ML             & CS & GP & GPT & ML            & CS & GP & GPT & ML             \\
\midrule
GCDC            & \textbf{0.50} & 0 & 0 & 0       & \textbf{0.50} & 0 & 0 & 0      & 0.43 & \textbf{1} & 0.67 & 0     \\
Meat sale       & \textbf{0.29} & \textbf{0.29} & 0.24 & 0.24       & \textbf{0.24} & \textbf{0.24} & \textbf{0.24} & \textbf{0.24}      & \textbf{1} & 0.86 & \textbf{1} & 0.34     \\
Pizza delivery  & \textbf{1} & 0.67 & 0.67 & 0.67       & 0.33 & \textbf{0.67} & \textbf{0.67} & 0.33   & 0.44 & \textbf{1} & 0.75 & 0.5     \\
Shipper-carrier & \textbf{1} & \textbf{1} & \textbf{1} & \textbf{1}       & \textbf{1} & \textbf{1} & \textbf{1} & \textbf{1}      & 0.39 & \textbf{1} & \textbf{1} & 0.75     \\
Transactive Energy & \textbf{0.04} & 0 & 0 & 0       & \textbf{0.04} & 0 & 0 & 0      & \ding{55} & 0 & 0 & 0     \\
\bottomrule
\end{tabularx}
\begin{tablenotes}
\item CS -- Claude 3.5 Sonnet. GP -- Gemini 1.5 Pro. GPT -- GPT-4o. ML -- Meta LLama 3.1 70b Instruct. \ding{55} -- Path explosion.
\end{tablenotes}
\end{threeparttable}
\vspace{-12px}
\end{table*}

\begin{figure}[tb!]
\begin{outerbox}{}
\vspace{-5pt}
\begin{footnotesize}
    Your task is to help create a Solidity smart contract based on a provided legal contract. The process will involve several steps where you will analyze different aspects of the legal contract, document key elements and necessary assumptions, and then, in the final step, generate the Solidity code. Each step will focus on a specific part of the contract to ensure clarity and manage complexity effectively.

    \tcbsubtitle{Step 1: Contract Parties, Powers, and Obligations}
    Analyze the provided legal contract. Identify and list: 
    \begin{enumerate}
        \item The parties involved. 
        \item Their powers and obligations.
    \end{enumerate}
    Document any assumptions or interpretations necessary for understanding these elements. Here is the legal contract:

    ``````
    \textit{Legal contract text goes here.}
    ''''''
    \tcbsubtitle{Step 2: Triggers and Timeframes}
    Based on the previously identified parties, powers and obligations, now focus on:
    \begin{enumerate}
        \item Events that trigger actions within the contract. 
        \item Timeframes and deadlines relevant to these events, powers, and obligations. Note any interpretations or assumptions related to these triggers and timeframes.
    \end{enumerate}
    \tcbsubtitle{Step 3: Penalties and Dispute Resolution}
    Continuing from the triggers and timeframes, identify:
    \begin{enumerate}
        \item Any penalties for non-compliance or breaches. 
        \item Dispute resolution mechanisms specified in the contract.
    \end{enumerate}
    List assumptions or clarifications needed for implementing these features in the smart contract.

    \tcbsubtitle{Step 4: Legal and Security Compliance}
    Before generating the code, ensure the smart contract meets legal and security standards: 
    \begin{enumerate}
        \item Discuss potential legal issues and how they can be addressed in the Solidity code. 
        \item Identify necessary security features and error handling mechanisms to make the smart contract robust and secure. 
    \end{enumerate}
    Document all assumptions and legal considerations.

    \tcbsubtitle{Step 5: Smart Contract Generation}
    Now, compile all the information and insights gathered from previous discussions into a concise and efficient Solidity smart contract. The code should:
    \begin{enumerate}
        \item Be structured clearly with defined sections for parties, powers, obligations, triggers, and dispute resolutions.
        \item Include essential comments that clarify sections and key functions for future reference.
        \item Ensure legal and technical compliance as discussed, focusing on functional accuracy and security features.
    \end{enumerate}
    Generate the smart contract code with minimal additional commentary, keeping it focused and lean. Provide brief inline comments only where necessary to explain complex logic or important compliance elements.
\vspace{-5pt}
\end{footnotesize}
\end{outerbox}
\vspace{-5pt}
\caption{LLM Prompts for Smart Contract Generation}
\label{fig:llm_final_prompt}
\vspace{-21pt}
\end{figure}

\subsubsection{Code Generation} Each model is prompted to generate Solidity smart contracts with the selected legal contracts. We looked for trivial errors such as missing Solidity contract definitions and whether the code would be able to execute upon deployment. If any issues are found, we repeated the code generation step up to three attempts. The code with least issues (in terms of executability and initial variable values) out of all attempts is taken as the model's output for modeling.

\subsection{Illustrative Analysis: GCDC Contract}
\subsubsection{PN Model Complexity} The GCDC legal contract PN serves as our baseline with 15 places and 5 transitions. The PNs derived from LLM-generated smart contracts show varying structural complexity: Claude Sonnet produced the most complex model (10 places, 6 transitions), followed by Gemini Pro and Llama 3.1 70b (both with 7 places, 4 transitions), while GPT-4o yielded the simplest model (6 places, 4 transitions). These variations provide an initial indication of how differently each LLM interpreted and implemented the contract requirements. 

\subsubsection{Behavioral Analysis} Comparing the RGs, there are 12 behaviors in the legal contract's RG, followed by 28 in Claude Sonnet, 9 in Gemini Pro, and 4 in Llama and 3 in GPT. Analyzing them, we observe:
\begin{itemize}
    \item Missing Behaviors: All LLM outputs except for Claude Sonnet lack pause functionality, which is a critical feature of the legal contract. This deviation affects FES and fitness scores. Gemini Pro implemented block/unblock functionality but not pausing or blacklisting. GPT-4o has freezing capabilities but lacks pausing and blacklisting. Llama 3.1 70b has no pause, block, blacklist, or freeze functionality.
    
    \item Extra Behaviors: All smart contracts included various additional behaviors, sometimes affecting legal compliance. 
    
    \item Path Analysis: Of the 12 unique paths in the legal contract's RG, only 6 are correctly implemented in smart contracts, and that too only by Claude Sonnet. The other models have fewer paths, indicating they do not fully capture the legal contract's behaviors.
\end{itemize}

\subsubsection{Compliance Metrics} Table~\ref{tab:legal_compliance_metrics} shows the compliance metrics for selected contracts. For GCDC, we observe:
\begin{itemize}
    \item Claude Sonnet has the highest fitness score at 0.5, denoting the best strict event equivalence. All other models score zero due to missing pause functionality, which is the terminating state of all behaviors from the legal contract. Without this functionality, LLM outputs do not reach the same final state as the legal contract.
    
    \item Claude Sonnet also has the best functional equivalence score of 0.5, indicating the best overall functional alignment with the legal contract. Again, the other models have a score of zero due to missing pause functionality.
    
    \item Gemini Pro has the highest precision score of one, indicating that all behaviors of its smart contract are present in the legal contract. Claude Sonnet scores 0.43, indicating that only 43\% of the behaviors in its smart contract are found in the legal contract. Llama 3.1 70b scores zero due to all of its behaviors being permitted to be executed by regular users, which is illegal as per the legal contract---smart contract actions must be initiated by authorized addresses.
\end{itemize}

\subsection{Summary of Quantitative Findings}
Across the 20 smart contracts generated by four LLMs for five legal contracts, we observe varying performance across our three compliance metrics, as seen in Table~\ref{tab:legal_compliance_metrics}.

For fitness, Claude 3.5 Sonnet achieves the highest scores on GCDC contract at 0.5, and Transactive Energy contract at 0.04, while performing comparably to other models on simpler contracts. All LLMs achieve identical fitness scores (0.24) on the Meat sale contract, suggesting similar capability in capturing its core behaviors. Similarly, the Shipper-carrier contract sees a perfect fitness score of 1 across all models.

For precision, Gemini 1.5 Pro generally achieves high scores, reaching 1 on GCDC, Pizza delivery and Shipper-carrier contracts. GPT-4o shows strong precision with scores between 0.67 and 1 on all but the Transactive Energy contract. Claude 3.5 Sonnet demonstrates lower precision scores (0.39-0.44) despite higher fitness, indicating generation of additional behaviors beyond the legal contract. Llama 3.1 70b shows variable precision performance, ranging from 0 to 0.75. All models show poor precision scores against the Transactive Energy contract.

The FES values are always equal to or higher than fitness, and they demonstrate a clear pattern based on contract complexity. The simpler Shipper-carrier contract achieves perfect FES of 1 across all models, while Pizza delivery shows consistent FES of 0.67 for all models except Claude 3.5 Sonnet, which achieves a perfect score. The Meat sale contract yields moderate FES values between 0.24-0.29 across all models. In contrast, the more complex Transactive Energy contract results in significantly lower scores, with Claude 3.5 Sonnet achieving only 0.04 and other models scoring 0.

\section{Discussion}
\label{sec:discussion}

\subsection{Comparison of LLM Outputs}

\subsubsection{Coding Patterns and Implementation Approaches}
We observe distinct patterns in how different LLMs approach smart contract generation. Claude 3.5 Sonnet, Gemini 1.5 Pro, and GPT-4o consistently use OpenZeppelin library contracts such as \texttt{Pausable} and \texttt{Ownable} that are widely used in the Ethereum development community. This results in more readable and maintainable code, as well as a higher likelihood of legal compliance. In contrast, Llama 3.1 70b tends toward self-implemented functionality, often producing error-prone partial implementations that lack robust security features.

While all models generate code targeting Solidity version 0.8.0 or higher, none fully leverage the language's newer features. For instance, all models continue to use the SafeMath library despite version 0.8.0's built-in overflow protection, suggesting their training data may not fully reflect current best practices.

\subsubsection{Performance Patterns and Contract Complexity Effects}
Our evaluation reveals a fundamental trade-off between precision and coverage in LLM-generated smart contracts. Models achieving higher precision scores, such as Gemini 1.5 Pro and GPT-4o, tend to generate more focused implementations but sometimes miss implementing complete contract functionality. Conversely, Claude 3.5 Sonnet achieves better functional coverage at the cost of introducing extraneous behaviors. FES values are at least equal to fitness in all instances, as the former relaxes the strict event equivalence condition of the latter.

The impact of contract complexity is particularly evident in our results. Simple contracts with fewer legal positions, such as the Shipper-carrier contract (3 obligations, no powers), see consistently high performance across all metrics and models. However, common failure scenarios emerge with more complex contracts:
\begin{itemize}
   \item Incomplete implementation of contracts' legal powers and obligations, especially noticeable in GCDC ToS (7 powers), where only Claude 3.5 Sonnet achieved non-zero fitness scores.
   \item Incorrect handling of administrative functionality such as pausing and access controls, notably in Llama 3.1 70b's implementations.
\end{itemize}

These patterns suggest that while LLMs can generate syntactically correct smart contracts, they struggle with the semantic complexity of legal requirements. The Transactive Energy contract (3 obligations, 3 powers) particularly demonstrates this challenge; it exhibited a path explosion problem with hundreds or even thousands of behaviors across LLM-generated smart contract RGs. Calculating the metrics, which requires pair-wise path comparisons, quickly becomes infeasible in such cases. We optimized the analysis by identifying subsequences of events with clear legality violations and discounting them from metric calculations. This speeds up the analysis, but does not address the combinatorial growth in path comparisons. In future, we plan to explore efficient methods for comparing RGs, e.g., using machine-learning techniques to identify common patterns and reduce the number of comparisons, or using LLMs to interpret individual paths and identify legal violations directly.

In summary, current LLMs are better suited for generating initial implementations of simpler contracts, requiring careful review and refinement for more complex scenarios.

\subsection{Analysis of Metrics with Properties of Software Measures}
\label{weyuker_analysis}
We evaluate the proposed metrics using Weyuker's informal properties for software measures~\cite{weyuker}, adapting them from their original purpose of evaluating syntactic complexity to our semantic evaluation of legal compliance. Using property names from Gustafon~\cite{gustafonSoftwareMeasures}, we find the \emph{distribution} property ($(\exists P)(\exists Q)$ such that $|P| \neq |Q|$ where $P, Q \in \mathcal{P}$, programs) requires different metric values for different programs, which our metrics satisfy as shown in Table~\ref{tab:legal_compliance_metrics}. \emph{Fineness} requires only finite programs to have any given metric value $c$, which holds true given practical implementation limits. \emph{Coarseness} ($ \exists$ distinct programs $P \text{, } Q$ such that $|P| = |Q|$) requires that distinct programs can have the same metric value, also demonstrated in Table~\ref{tab:legal_compliance_metrics}. 
The remaining Weyuker's properties primarily concern program syntax and are thus less relevant for our semantic compliance measures.

Examining formal properties of software measurement~\cite{gustafonSoftwareMeasures}, we find our metrics are built on \emph{non-trivial} PN models with multiple possible instances and exhibit \emph{non-trivial ordering}. All three metrics have a ratio scale, enabling meaningful comparisons of relative values (e.g., one program having double the fitness of another). Therefore, these properties enable the metrics to  effectively compare smart contracts, making them useful for choosing between LLMs for code generation.

\subsection{Threats to Validity}
\label{sec:threats}

\textbf{Construct validity}. The accuracy of the proposed metrics depends on the correct interpretation of legal text, which remains challenging. We mitigated this by independently verifying our contract interpretations with two lawyers and using tools such as CPN IDE to simulate PN models and catch inconsistencies. Some legal nuances might still be lost in the translation to PN representations.

\textbf{Internal validity}. When generating smart contracts, variations in prompts could affect the quality of the output. We standardized prompts across all models and tried multiple prompt variants to mitigate this impact.

\textbf{External validity}. While we studied various contract types, LLMs, and smart contracts, we tested only specific subsets of terms in some contracts, as noted with USDC Terms of Service, to manage complexity. Additional experiments with complete contracts and a wider range of legal documents would further validate generalizability of proposed metrics.

\textbf{Conclusion validity}. The lack of comparable frameworks for nuanced legal compliance evaluation makes it difficult to directly benchmark our metrics. Alternatively, we analyze the metrics against syntactic software quality properties.

\vspace{-2pt}
\section{Related Work}
\label{sec:literature}

We review previous work in three key areas: LLM legal reasoning and code generation, formal and mixed methods for smart contract testing and verification, and model-based approaches for legal compliance.

\textbf{LLM legal reasoning and code generation.} Chen et al.\cite{chen2021evaluating} conducted extensive evaluations of LLM-generated code quality, finding that while models can generate syntactically correct code, semantic correctness remains challenging. For smart contracts specifically, Barbara et al.\cite{barbara2024automatic} demonstrated that a contemporary LLM output fell short of generating production-ready smart contracts, but did not provide a method to quantify compliance. LegalBench, a benchmark for natural-language legal reasoning tasks~\cite{guha2023legalbench}, notes that LLMs are particularly challenged by legal-rule application and conclusion tasks, but does not evaluate coding capabilities of LLMs.

\textbf{Formal and mixed methods}. Previous works have focused on vulnerability detection through symbolic and concolic execution, with tools such as Manticore~\cite{manticore} and Oyente~\cite{oyente}.  Neither legal compliance nor its measurement are explicit goals of these tools, while they may be indirectly useful via functional correctness validation. The Symboleo framework\cite{sharifi2020symboleo} made significant advances in modeling contract behavior. Their conformance checker SymboleoCC~\cite{parvizimosaed2020subcontracting} points to the feasibility of automated compliance verification, though limited to contracts written in their specification language and manually written test cases. Logic-based languages~\cite{idelberger2016evaluation, governatori2018legal} and Domain-Specific Languages~\cite{adico, spesc} also represent contract elements without quantifying compliance.

\textbf{Model-based approaches.} Previous approaches have sought to bridge legal and programming domains~\cite{amato_model_law_compliance_sc}, but lack systematic compliance quantification. Ladleif et al.\cite{ladleif2019unifying} developed a UML meta-model for legal concepts in smart contracts, while Symboleo\cite{sharifi2020symboleo, parvizimosaed2020subcontracting} provides a specification language for modeling contract lifecycles and legal positions. However, these approaches stop short of providing comprehensive compliance evaluation mechanisms.

Our approach differs from prior work in three key aspects. First, while existing LLM studies evaluate either code quality or legal reasoning in isolation, we bridge these aspects with quantitative metrics to measure legal compliance of smart contracts. Second, unlike the custom specification requirements of tools such as SymboleoCC, our PN-based process modeling approach can generically model and compare any smart contract against a legal contract. Finally, while previous modeling approaches lack systematic compliance measurement, our behavior-based metrics suite provides a comprehensive framework for quantifying alignment between legal intentions and technical implementations. This enables granular compliance assessment and comparison of different LLMs' capabilities in generating legally compliant smart contracts.

\section{Conclusion}
\label{sec:conclusion}

Our evaluation of 20 smart contracts from four leading LLMs reveals that they can generate syntactically valid code but vary in their legal compliance capabilities. While Claude Sonnet offers better functional coverage with excess behaviors, Gemini Pro and GPT-4o prioritize precision over completeness. Our behavior-based measurement approach and resulting metrics are applicable to both LLM-generated and manually written code, and the PN-based analysis framework enables formal compliance evaluation. The proposed metrics demonstrate that while LLMs show promise in generating legally compliant smart contracts, they face challenges in consistently capturing the full scope of legal requirements and avoiding extraneous behaviors. These findings highlight the need for careful human review prior to deploying LLM-generated smart contracts. Future work could explore the relationship between LLM architectures, training processes, and their ability to maintain legal compliance, automate compliance checking through our metrics framework, and address the state space explosion challenges we observed in complex contracts.

\section*{Acknowledgement}
We gratefully acknowledge funding from the Digital Finance CRC which is supported by the Cooperative Research Centres program, an Australian Government initiative. We also thank the Australia and New Zealand Bank for their support and guidance during this research.

\bibliographystyle{ieeetr}
\bibliography{paper}

\end{document}